%% file: Linden_feast_cigale_submm_v2.tex
\documentclass[twocolumn]{aastex631}

\usepackage{amsmath}
\usepackage{array}   
\usepackage{enumitem}
\usepackage{hyperref}

\begin{document}

\title{Feedback in Extragalactic Star Clusters (FEAST): Spectral Energy Distributions and the Physical Properties of Star Clusters in NGC 628 with CIGALE}

\correspondingauthor{S. T. Linden}
\email{seanlinden@arizona.edu}

\author[0000-0002-1000-6081]{Sean T. Linden}
\affiliation{Steward Observatory, University of Arizona, 933 N Cherry Avenue, Tucson, AZ 85721, USA}

\author[0000-0002-8192-8091]{Angela Adamo}
\affiliation{Department of Astronomy, The Oskar Klein Centre, Stockholm University, AlbaNova, SE-10691 Stockholm, Sweden}

\author[0000-0002-8222-8986]{Alex Pedrini}
\affiliation{Department of Astronomy, The Oskar Klein Centre, Stockholm University, AlbaNova, SE-10691 Stockholm, Sweden}

\author[0000-0002-5189-8004]{Daniela Calzetti}
\affiliation{Department of Astronomy, University of Massachusetts Amherst, 710 North Pleasant Street, Amherst, MA 01003, USA}

\author[0000-0002-2199-0977]{Helena Faustino Vieira}
\affiliation{Department of Astronomy, The Oskar Klein Centre, Stockholm University, AlbaNova, SE-10691 Stockholm, Sweden}

\author[0000-0002-1723-6330]{Bruce G. Elmegreen}
\affiliation{Katonah, NY 10536, USA}

\author[0000-0001-8608-0408]{John S. Gallagher}
\affiliation{Department of Astronomy, University of Wisconsin-Madison, 475 N. Charter Street, Madison, WI 53706, USA}

\author[0000-0001-8348-2671]{Kelsey Johnson}
\affiliation{Department of Astronomy, University of Virginia, Charlottesville, VA, USA}

\author[0000-0003-1427-2456]{Matteo Messa}
\affiliation{INAF - Osservatorio di Astrofisica e Scienza dello Spazio di Bologna, Via Gobetti 93/3, I-40129 Bologna, Italy}

\author[0000-0002-3247-5321]{Kathryn~Grasha}
\affiliation{Research School of Astronomy and Astrophysics, Australian National University, Canberra, ACT 2611, Australia}

\author[0000-0002-5259-4774]{Ana Duarte-Cabral}
\affiliation{Cardiff Hub for Astrophysics Research and Technology (CHART), School of Physics \& Astronomy, Cardiff University, The Parade, CF24 3AA Cardiff, UK}

\author[0000-0001-8068-0891]{Arjan Bik}
\affiliation{Department of Astronomy, The Oskar Klein Centre, Stockholm University, AlbaNova, SE-10691 Stockholm, Sweden}

\author[0009-0003-6182-8928]{Giacomo Bortolini}
\affiliation{Department of Astronomy, Oskar Klein Centre, Stockholm University, AlbaNova University Centre, SE-106 91 Stockholm, Sweden}

\author[0000-0001-6464-3257]{Matteo Correnti}
\affiliation{INAF Osservatorio Astronomico di Roma, Via Frascati 33, 00078, Monteporzio Catone, Rome, Italy}
\affiliation{ASI-Space Science Data Center, Via del Politecnico, I-00133, Rome, Italy}

\author[0009-0009-5509-4706]{Drew Lapeer}
\affiliation{Department of Astronomy, University of Massachusetts, 710 North Pleasant Street, Amherst, MA 01003, USA}

\author[0000-0001-8490-6632]{Thomas S.-Y. Lai}
\affiliation{IPAC, California Institute of Technology, 1200 E. California Blvd., Pasadena, CA 91125, USA}

\author[0000-0002-3005-1349]{G\"oran \"Ostlin}
\affiliation{Department of Astronomy, Oskar Klein Centre, Stockholm University, AlbaNova University Centre, SE-106 91 Stockholm, Sweden}

\author[0000-0002-0806-168X]{Linda J. Smith}
\affiliation{Space Telescope Science Institute, 3700 San Martin Drive, Baltimore, MD, 21218, USA}

\author[0000-0002-0986-4759]{Monica Tosi}
\affiliation{INAF - Osservatorio di Astrofisica e Scienza dello Spazio di Bologna, Via Gobetti 93/3, I-40129 Bologna, Italy}

\begin{abstract}
With Hubble Space Telescope (HST) and James Webb Space Telescope (JWST) observations of NGC~628 spanning 0.3--7.7\,$\mu$m, we fit the spectral energy distributions (SEDs) of over 12,000 optically-selected star clusters, emerging young star clusters (eYSCs), and MIRI-selected sources with \textsc{cigale} to derive their ages, masses, extinctions, and dust properties. We find that near-infrared selected eYSC-I (compact Pa$\alpha$ and 3.3,$\mu$m PAH emission) and eYSC-II (compact Pa$\alpha$ and diffuse 3.3,$\mu$m PAH emssion) sources peak at $\sim$3--5~Myr, where $\sim 12\%$ of the clusters have an $E(B{-}V)>2$, demonstrating the presence of dust-embedded populations as clusters emerge. Further, the distributions of the fractional polycyclic aromatic hydrocarbon (PAH) abundance ($q_{\mathrm PAH}$) and stellar-to-nebular attenuation ratio ($E(B{-}V)_{{\rm \star}}/E(B{-}V)_{\rm neb}$) suggest an evolutionary sequence in which sources evolve from eYSC-I to eYSC-II as clusters clear their surrounding dust and gas. The photo-dissociation region (PDR) clearing timescale inferred from the ratio of eYSC-I to optically visible stellar clusters is $\sim$4~Myr. Additionally, we find that star clusters in the spiral arms of NGC 628 are preferentially more massive and more dust-reddened than those in inter-arm regions.~Finally, we find that $\sim$65\% of eYSC-I, $\sim$27\% of eYSC-II, and $\sim$40\% of F335M-selected sources coincide with an F770W peak in our MIRI-selected catalog within 4 pixels, confirming that F770W-bright sources preferentially trace the youngest and dustiest regions. Overall, our results highlight the ability of JWST together with \textsc{cigale} model grids to identify and characterize eYSCs during their short-lived embedded phases, and provide constraints on the feedback mechanisms that govern the emergence of stellar clusters.
\end{abstract}

\keywords{galaxies: star clusters - galaxies: ISM - galaxies: NGC 628}

\section{Introduction}

The complex interplay between star formation and stellar feedback at parsec scales has now been established as one of the key factors that regulate galaxy growth across cosmic time \citep{schinnerer24}. Young stellar clusters (YSCs) with $R_{eff} \sim 3$ pc \citep{bag21} form in dense molecular gas clumps as a result of the hierarchical fragmentation of cold gas \citep[e.g.,][]{elmegreen04}, and host the majority ($>80\%$) of the massive stars responsible for stellar feedback \citep{ll03,soey17}, with embedded super star clusters in particular tracing the most extreme, compact star-forming environments \citep{kj99,kej01,kej04}. Further, YSCs are observed to be ubiquitous in both nearby \citep[][]{krumholz19} and high-redshift \citep{vanzella21, vanzella23, claeyssens23, aa24, messa25, claeyssens25, claeyssens26} star-forming galaxies. Yet they remain a challenge for both observations and simulations, which struggle to self-consistently follow star and cluster formation at the required physical resolution \citep{grudic22}. 

JWST has opened a new window into the ``emerging phase" of newly formed stars and star clusters (i.e., emerging YSCs: eYSCs). Numerical simulations suggest that during these phases photo-ionization and mechanical feedback from stellar winds and supernovae (SNe) produced by massive stars ($\geq 10$ $M_{\odot}$) lead to the dissolution of the natal giant molecular clouds (GMCs) where they formed \citep[][]{wall20,guszejnov22,lahen23,deng24}. However, important aspects of the feedback process, such as the timescales and conditions necessary for embedded YSCs to emerge from their parent GMCs, are still unknown observationally \citep[e.g.,][]{sokal15,sokal16}. This emergence process is critical because it directly affects the density and temperature of the interstellar medium (ISM), helping to regulate the local efficiency of star formation \citep{ma20,grudic21} and the overall recycling of gas \citep{bending20, hopkins20}. Initial JWST studies reported the detections of eYSCs across both the discs and nuclear regions of starburst galaxies \citep{bcw23,linden23,linden24}. These eYSCs show signatures of hot dust emission in their near-IR (NIR) colors and bridge the gap between GMCs and YSCs. Increasing evidence points toward an ``optically-dark'' embedded phase that lasts less than 5 Myr \citep{kim23,kim25,ramambason26}. The detection of these eYSCs opens an unprecedented opportunity to determine the rapidly changing physical properties of the gas and dust exposed to the feedback of newly formed stars.

However, the measurement of the physical properties of emerging stellar populations using broad- and narrow-band photometry is an arduous task with multiple physical phenomena involved: as starlight interacts with dust, a fraction is absorbed and then re-emitted at longer wavelengths and leads to extinction and reddening of the stellar emission. Further, eYSCs contain massive stars that ionize the surrounding gas, leading both to the rise of a nebular continuum and to the appearance of a series of emission lines. This nebular emission can be quite bright, and may represent a non-negligible fraction of the flux captured by broadband filters \citep[e.g.,][]{groves08, reines09}. To further our understanding of eYSC formation and evolution, it is necessary to carefully account for these effects to accurately measure cluster physical properties (age, mass, and extinction).

Over the past few decades, various codes have been developed to model the panchromatic emission from galaxies. Using various fitting techniques including $\chi_{\rm red}^{2}$ minimization and Bayesian inference to constrain physical parameters, these codes can provide deeper insight into the properties of star cluster populations \citep[e.g.,][]{dacunha08,moustakas13,cc16}.The Legacy ExtraGalactic Ultraviolet Survey \citep[LEGUS;][]{calzetti15} utilized a simple stellar population model \citep[e.g., Yggdrasil;][]{Yggdrasil} to derive ages, masses, and reddening for optically-visible star clusters, further advancing our understanding of cluster formation and evolution \citep[e.g.,][]{aa17,mm18a,mm18b,aa20a,linden22a,mcquaid24,deshmukh24}. Here, we report on the augmentation of the publicly available SED fitting package \textsc{cigale}: Code Investigating GALaxy Emission \citep{cigale} to fit eYSC populations in nearby galaxies. This tool allows for flexibility in modeling stellar evolution tracks, star formation histories, dust attenuation and emission, and nebular emission, making it well-suited for studying stellar populations from the ultraviolet (UV) to the NIR \citep{turner21}.

The Feedback in Extragalactic Star Clusters (FEAST) survey uniquely combines high-resolution, multi-wavelength imaging from HST and JWST to probe the earliest phases of star cluster evolution on parsec scales. FEAST is specifically designed to capture the embedded and emerging phases of star clusters by jointly tracing stellar emission, ionized gas, and dust-reprocessed light from the UV through the mid-infrared. This approach enables us to directly connect the physical properties of young clusters to their surrounding interstellar medium, providing new constraints on feedback-driven cloud clearing, dust processing, and the timescales governing cluster emergence that are otherwise inaccessible with previous UV-optical datasets. Overall, we present an analysis of the physical properties of star clusters in using \textsc{cigale} model-fitting to shed light on the fundamental processes governing star formation and feedback in NGC 628. 

The paper is organized as follows: In \S 2, the observations, cluster identification and photometry are presented. In \S 3 we outline our spectral energy distribution model fitting framework with \textsc{cigale}. In \S 4, we analyze the resulting best-fit age, mass, and extinction distributions for eYSCs, and compare our results to other studies of emerging and embedded star clusters in nearby galaxies in \S 5. In \S 6 we summarize the results.

\begin{figure*}[!htb]
\centerline{\includegraphics[width=\textwidth]{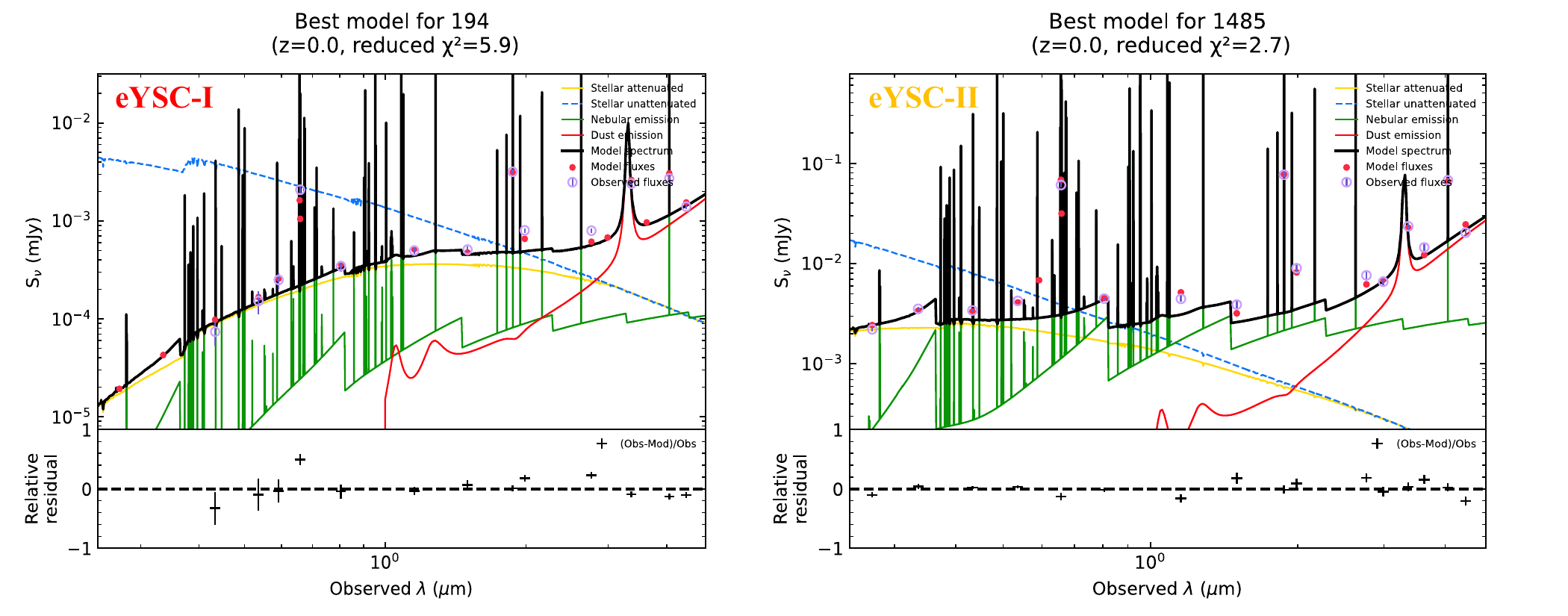}}
\caption{Examples of spectral energy distribution (SED) fits for representative eYSC-I (\textbf{Left}) and eYSC-II (\textbf{Right}) sources identified in NGC 628. The solid black line shows the best-fitting \textsc{cigale} model, with the unattenuated (attenuated) stellar emission shown in blue (yellow), nebular emission in green, and the dust emission in red. The observed and model fluxes corresponds to the purple and red points respectively. These examples illustrate an evolutionary sequence where eYSC-I sources  are very young, dust-embedded clusters with strong NIR excess and nebular emission, and eYSC-II sources are moderately young clusters where extinction is lower and PAH/dust features are weaker.}
\end{figure*}

\section{Observations, Source Selection, and Photometry}

\subsection{Observations}
JWST NIRCam and MIRI observations of NGC 628 have been obtained as part of the FEAST program GO1783 (PI: A. Adamo). For NIRCam we obtained simultaneous observations in the short wavelength (F115W, F150W, F187N, F200W) and long wavelength channels (F277W, F335M, F405N, F444W). We employ a FULLBOX 4TIGHT dither pattern to ensure high quality sampling of the point spread functions (PSFs) over a large field of view (2' x 6'). The MIRI observations have been obtained in the F560W and F770W filters. Five MIRI pointings were necessary to cover the same area of the NIRCam mosaic. Additionally, for the MIRI observations, an external sky background pointing was included for calibration.

We refer to Adamo et al., (in prep.) for an in-depth description of the data processing, while here we present an overview of the main steps. We reduced the data using the distributed pipeline version 1.12.5 with calibration data context number 1174 for NIRCam and version 1.11.4 using context number 1141 for MIRI. For the MIRI observations, we replaced the default SkyMatchStep of the JWST level 3 pipeline with the PixelSkyMatchStep \citep{VarunSoftware}. This change is implemented to remove the strong gradients in the background of the final mosaic for MIRI data caused by computing the differences between medians of pixels in overlapping regions of pairs of exposures. Further, we download additional HST WFC3/UVIS and WFC/ACS archival observations (F275W, F336W, F435W, F555W, F657N, F658N, F606W, and F814W) covering the UV-optical range from the Mikulski Archive for Space Telescopes (MAST). The final mosaics for all 18 filters are aligned to the same reference system using GAIA \citep{gaiadr2}. Finally, we resampled our aligned mosaics to a common resolution of 0.04''/pix for HST and NIRCam and 0.08''/pix for MIRI respectively.

\input{catalog_numbers.tex}
\input{table_parameters_class123.tex}
\input{table_parameters_optical.tex}

\subsection{Catalog extraction}

\subsubsection{eYSC catalogs}
In this study we make use of catalogs of young and emerging star clusters in NGC~628 \citep[presented in][]{pedrini24, pedrini25, gregg24}.~Candidate eYSCs are extracted from the continuum-subtracted maps tracing the Pa$\alpha$ (F187N), Br$\alpha$ (F405N), and 3.3 $\mu$m PAH emission \citep[see][for a description of the continuum subtraction procedure]{gregg24}. Source detection is carried out independently on the three emission line maps using the \texttt{FEAST-pipeline} (Adamo et al. in prep) with the \texttt{SEP} package \citep{sep}, a Python implementation of \texttt{SExtractor} \citep{sex}, resulting in three independent catalogs. For initial detection we retain all sources with a $S/N > 3$ in the F187N, F200W, F335M, and F444W filters. All extracted sources were then visually inspected using cutouts of both the multi-band HST and JWST imaging as well as the continuum-subtracted emission line maps to assess their morphology and the local environment surrounding each cluster. Only sources with compact morphologies consistent with star cluster candidates were retained for subsequent analysis. During this inspection, contaminants such as foreground stars, background galaxies, diffraction spikes, detector artifacts, and blended or morphologically irregular sources were identified and removed from the final catalog.

As a final step, we matched the three catalogs by their positions to determine their classification: The eYSC-I catalog consists of clusters that display compact 3.3\,$\mu$m PAH emission together with peaked Pa$\alpha$ emission, identifying the most deeply embedded systems where dust and ionized gas remain closely associated with massive stars. The eYSC-II catalog includes clusters with compact Pa$\alpha$ emission but lacking a similar compact counterpart in 3.3\,$\mu$m emission, representing sources that are slightly more evolved, with PAH emission already diminished or partially cleared. The F335M-selected catalog is defined having sources with a strong compact 3.3\,$\mu$m PAH feature, that do not have a similar counterpart in Pa$\alpha$ emission. The total number of eYSC-I, II, and F335M-selected sources extracted for SED fitting is 2213 with 835 eYSC-I, 741 eYSC-II, and 637 F335M-selected sources in each class respectively.

\subsubsection{MIRI catalog}

Observations with MIRI are sensitive to longer-wavelength PAH emission and warm dust continuum, allowing us to probe the youngest and dustiest phases of emerging star clusters in nearby galaxies. Using the concentration index (CI, defined as the magnitude difference between a $1$-pixel and $3$-pixel radius aperture) we identify compact ($1 \lesssim {\rm CI} \lesssim 3$) F770W candidates using the \texttt{SEP} package and produce a MIRI-selected catalog of sources with a $S/N > 3$ in both the F560W and F770W filters. This results in an initial sample of 1328 F770W-bright sources. However, compact emission in the F770W band can originate not only from embedded star clusters but also individual evolved stars. Therefore, to minimize contamination, the final F770W catalog only retains sources that overlap within $0.24\arcsec$ (3 MIRI pixels) with at least one of the eYSC candidates extracted from the NIRCam data, resulting in a final catalog of 1023 sources. The overlap fractions are high for the youngest and dustiest populations: $\sim$65\% of eYSC-I, $\sim$27\% of eYSC-II, and $\sim$40\% of F335M-selected sources coincide with an F770W peak. 

\subsubsection{Optical YSC Catalogs}

Finally, using the HST/F555W filter as our reference image we implement the \texttt{SEP} package to produce an optically-identified (oYSC) catalog for NGC 628. Candidate oYSC sources are required to satisfy the following criteria: (1) a measured effective radius consistent with $R_{\rm eff} \lesssim 10$~pc at the distance of NGC~628 (corresponding to $\lesssim 0\farcs21$), and (2) a CI within the range $1.3 \lesssim {\rm CI} \lesssim 1.8$~mag. The lower bound excludes unresolved point sources such as stars, while the upper bound rejects diffuse associations and background galaxies not removed with our first cut \citep{aa17,mm18a,mm18b}. These cuts ensure that the resulting catalogs are dominated by compact stellar clusters suitable for SED fitting. These catalogs are then visually inspected to minimize spurious detections in regions of bright diffuse emission. From this initial catalog of 9197 detections within the JWST NIRCam FOV we additionally selected clusters with an assigned morphological class of 1, 2 or 3 \citep[for a description of these classes see][]{aa17}, and a fitted age of $< 10$ Myr with \textit{Yggdrasil} SSP models to produce a final sample of 1329 optically-selected oYSCs to be fit with \textsc{cigale}.

\subsection{Photometry}

The positions of the confirmed eYSC and oYSC candidates in the NIRCam, MIRI, HST imaging are used by the \texttt{FEAST-pipeline} to provide a uniform framework for photometry, and catalog assembly across the FEAST sample. Photometry is measured in consistent circular apertures of $0.16"=4$ pix, a sky annulus at 5-7 pix, and aperture correction up to 20 pix across all HST and JWST NIRCam bands after images are aligned to a common WCS grid and PSFs are homogenized to the resolution of the longest-wavelength NIRCam filter. The aperture corrections are calculated as $0.8"$ (20 NIRCam pixels). For our MIRI imaging where the pixel scale is 2x larger, photometry is performed in circular apertures of $0.4"=5$ pix and a sky annulus at 6-7 pix. We then match the aperture in arcsec across all other HST and NIRCam bands which have all been PSF matched to the F770W filter. Aperture corrections in the MIRI catalog are done up to $0.8"$ (similar to the previous catalogs). Non-detections in all filters are handled as upper-limits in \textsc{cigale}. A final cut is applied to the sample to retain a total of 1942 sources (684 eYSC-I, 651 eYSC-II, and 607 F335M-selected) that are well-fit ($\chi_{\rm red}^{2}<20$) by \textsc{cigale} (see Section 3). In the following analysis we refer to 'eYSC catalogs' as the combination of eYSC-I, eYSC-II, and F335M-selected sources. A summary of the detection statistics and final catalogs is given in Table 1.

\begin{figure*}[!htb]
\centerline{\includegraphics[width=\textwidth]{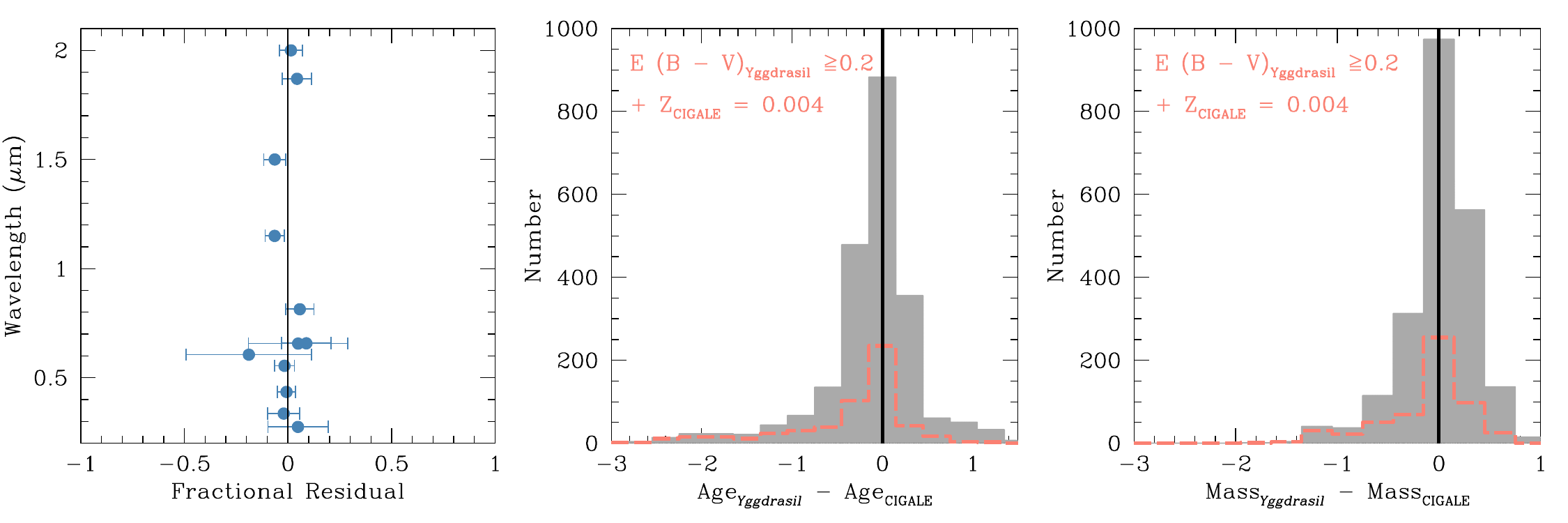}}
\caption{\textbf{Left:} The fractional residuals between our observed optically-selected star cluster catalog and the best-fitting model fluxes based on \textsc{cigale} when restricted to $\lambda < 2$\,$\mu$m to better-match SSP models which do not include any contributions from hot dust or PAH emission. \textbf{Middle:} Histograms of the difference in the derived age of optically-selected star clusters in NGC 628 relative to \textit{Yggdrasil} stellar population fits. \textbf{Right:} Histograms of the difference in the derived mass relative to \textit{Yggdrasil} stellar population fits. In both panels the distribution of sources with a best fit metallically of z$=0.004$ and an $E(B{-}V) \geq 0.2$ are shown in pink. Overall, we find that \textsc{cigale} can successfully fit the sample of optically-selected sources in NGC 628, and is in general agreement with the results of more traditionally-adopted SSP models.}
\end{figure*}

\section{\textsc{cigale}: Setup and Basic Considerations}

\textsc{cigale} operates by generating a grid of models based on the user's input parameters. We implement a simple stellar population (SSP) model by utilizing the double-exponential SFH module \texttt{sfh2exp} and inputting a very short e-folding time for the stellar population and a zero mass fraction of the second/late starburst population. \textsc{cigale} then compares the cluster's photometry with each model of the grid and calculates the $\chi_{\rm red}^{2}$ value to determine the goodness-of-fit. The $\chi_{\rm red}^{2}$ value is converted into a likelihood via $\exp(-\chi_{\rm red}^{2}/2)$. Once each model has been tested, \textsc{cigale} estimates the best-fit parameters in two ways: simple $\chi_{\rm red}^{2}$ minimization and a likelihood-weighted mean. $1\sigma$ uncertainties can be calculated by the difference between the best-fitting model and the models with $\chi_{\rm red}^{2}$ values of $1 + \chi_{\rm red, min}^{2}$. The likelihood-weighted mean of all models in the grid is computed and used as the Bayesian estimate of the physical properties, with $1\sigma$ uncertainties determined by the likelihood-weighted standard deviation across the full grid.

Although cluster mass is an output of the SED modeling, it is not treated as a third dimension of the model grid. The masses corresponding to a particular model on the age-reddening grid are determined directly from the chosen IMF and star formation history, and the mass is appropriately scaled based on the cluster's luminosity, with a fully-sampled IMF assumed. The effect of a stochastically sampled IMF is discussed in Section 5.6. 

For the eYSC and MIRI-selected sources, the age grid has ten linearly spaced models from 1 to 10~Myr ($\Delta T = 1$~Myr, the highest precision available to \textsc{cigale}). The upper age limit of 10 Myr is motivated by the selection of eYSC sources via detectable Pa~$\alpha$ and Br~$\alpha$ emission, which traces ionizing massive stars and declines rapidly on timescales of $\lesssim 10$ Myr \citep[e.g.,][]{ks98,dc10,ejm11b}. We allow the internal reddening to range from $E(B{-}V) = 0$ to $5$ mag in steps of 0.1~mag. Each model on the grid has a corresponding mass based on the \citet{bc03} SSP track and a fully sampled \citet{chabrier03} IMF. We assume solar metallicity, an escape fraction ranging from 0.01–0.6 \citep[e.g.,][]{izotov16,flury22}, and two values ($\delta = -0.25, 0$) for the slope of the UV–optical attenuation curve \citep[e.g.,][]{reddy20,shivaei20}. We further adopt a simplified set of assumptions for the dust emission model implemented in \textsc{cigale}, including three values of $\gamma$ to represent the fraction of dust heated by the young stellar population versus the diffuse interstellar radiation field, a single value for the near- to far-IR dust power-law slope, and three values of $U_{\min}$ \citep{dl07,draine14}. These choices are motivated by observations of H\textsc{ii} regions and star-forming galaxies in the nearby universe \citep{leja17,chastenet19,aniano20,chastenet25}. Finally, we adopt the full range of $q_{\mathrm PAH}$ values available in the model to explore the impact of PAH abundance variations on the resulting SEDs \citep{lin20,sutter24,shivaei24}. A more detailed explanation of each model parameter is given in Table 2.

For the optically-selected sources the age model grid consists of ten linearly-spaced models from 1~Myr to 10~Myr with $\Delta T=$~1~Myr and 100 log-spaced models from 11~Myr to 13~Gyr with $\Delta\log(T/{\rm Myr})\approx0.3$. The reddening $E(B{-}V)$ model grid spans from 0~mag reddening to 1.5~mag with $\Delta E(B{-}V) = 0.01$~mag. Each model on the grid has a corresponding mass based on the \citet{bc03} SSP track and the fully-sampled \citet{chabrier03} IMF. We assume both solar and sub-solar metallicities and an escape fraction which ranges from 0.01-0.6 \citep[e.g.,][]{izotov16,flury22}. The remaining model choices for our implementation of the \citet{draine14} model are the same as those adopted for our eYSC and MIRI-selected sources. The full grid of parameters adopted for eYSC, MIRI-selected, and optically-selected clusters in NGC 628 are listed in Tables 2 and 3.

The adopted model grid has since been applied successfully to SED catalogs of eYSCs in additional FEAST targets, including M83, M51, and NGC 4449. Specifically, \citet{knutas25} used these grid assumptions to model JWST NIRCam observations of emerging clusters in M83 and derived an average clearing timescale of $\sim 6$ Myr, with shorter emergence timescales ($\sim 5$ Myr) for more massive ($M\gtrsim5\times10^3$ M$_{\odot}$) clusters. Similarly, \citet{pedrini25} applied the same framework to the eYSC catalogs in NGC 628, M51, M83, and NGC 4449, identifying systematic NIR excesses at 1.5–2.5 $\mu$m associated with very young and low-mass clusters, pointing toward shortcomings in current dust and stellar models at these wavelengths. This latter study emphasizes the importance of including models for young stellar objects (YSOs) and to account for stochastic IMF sampling effects in low-mass clusters (e.g., $M\lesssim3\times10^3$ M$_{\odot}$), as the standard SED models sometimes assign older ages to clusters with strong Pa$\alpha$ emission. For a further discussion of these issues see Section 5.6.

In Figure~1 we show the results of \textsc{cigale} fitting for a representative star cluster from each of our eYSC-I (left Panel) and eYSC-II (right Panel) catalogs. In general, we find that eYSC-I sources have relatively red SEDs which rise from UV-NIR wavelengths, with strong hydrogen recombination line and 3.3$\mu$m PAH emission. However, eYSC-II sources appear to have flatter UV-NIR SEDs with weaker 3.3$\mu$m PAH emission suggesting an evolutionary sequence between the two classes. In the following Section we will explore the resulting distributions of derived physical properties for each of our star cluster catalogs in NGC 628.

\begin{figure*}[]
\centerline{\includegraphics[]{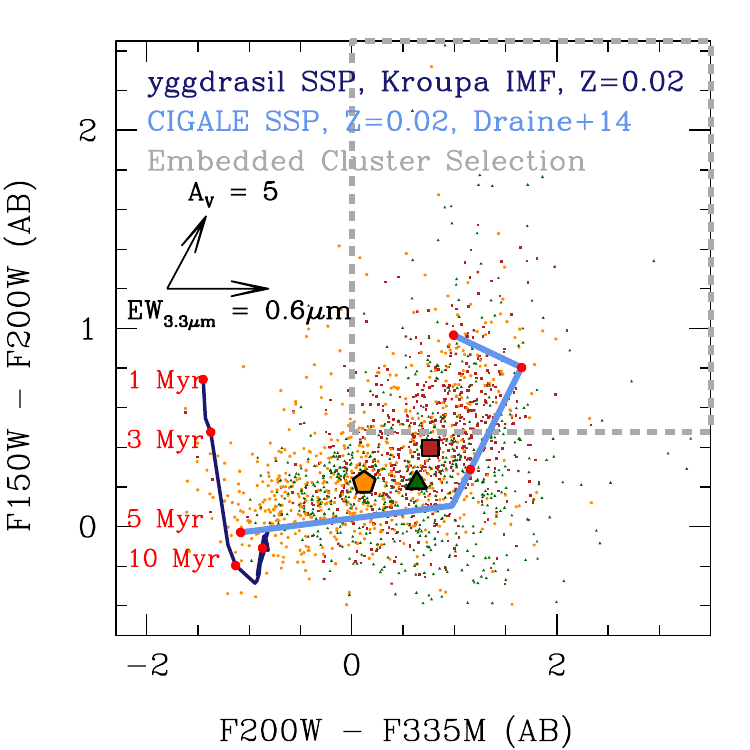}}
\caption{The near-IR color--color diagram of eYSC-I (red), eYSC-II (orange), and F335M-selected (green) sources in NGC~628 using F150W{-}F200W and F200W{-}F335M colors with the median value for each distribution shown as a square, pentagon, and triangle respectively. Overlaid in dark blue is a \texttt{Yggdrasil} SSP model track with solar metallicity and a Kroupa IMF \citep{pk01}. In light blue we show the same SSP model track using the \textsc{cigale} grid presented in Table 2 with the \citet{draine14} dust model.~The top-left arrows represent an extinction of $A_{V} = 5$ and the maximum contribution to the F335M flux from 3.3$\mu$m PAH emission determined for star-forming regions within galaxies observed as part of the PHANGS-JWST survey respectively \citep{sandstrom23}. This comparison demonstrates that purely stellar models under-predict the red F200W{-}F335M colors of the youngest sources, reflecting excess emission from dust and 3.3$\mu$m PAH emission for the young and most dust-embedded eYSCs in NGC 628. The gray box indicates the embedded-cluster color selection introduced by \citet{linden23,linden24}.}
\end{figure*}

\begin{figure*}[!htb]
\centerline{\includegraphics[width=\textwidth]{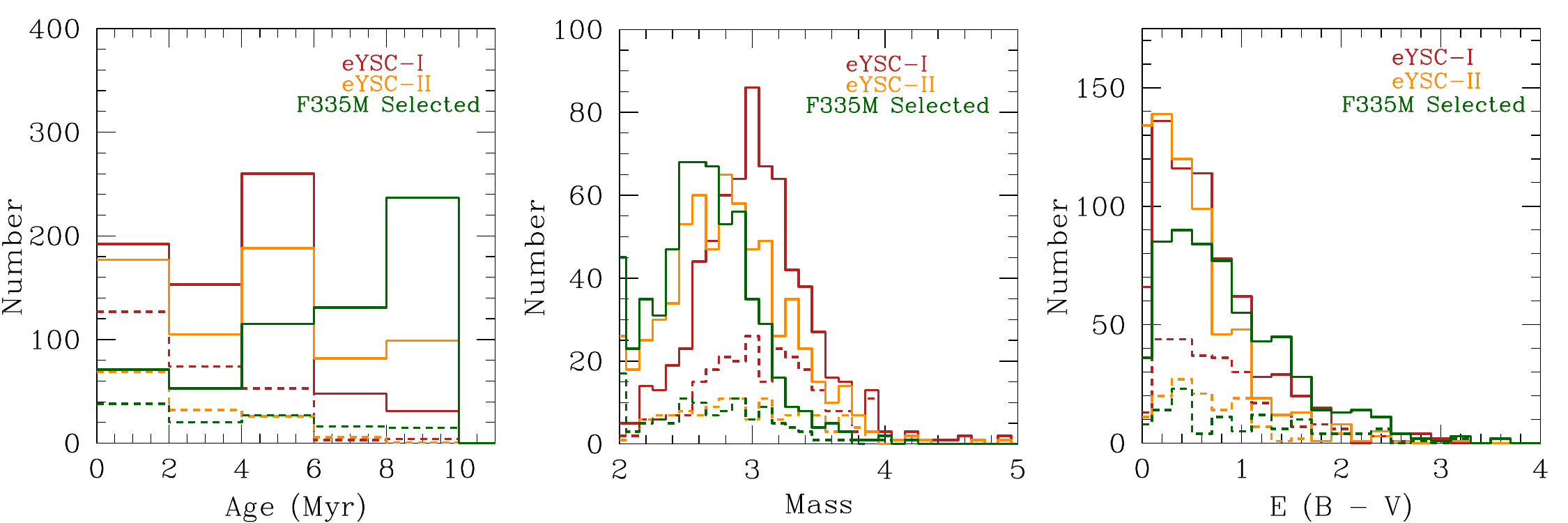}}
\caption{Results of the full \textsc{cigale} SED fitting for eYSC candidates in NGC~628. Panels show the distributions of best-fit ages (\textbf{Left}), stellar masses (\textbf{Middle}), and extinction $E(B{-}V)$ (\textbf{Right}). The distribution of ages for eYSC-I and II sources peaks at $\sim 3-5$ Myr. eYSC sources in NGC 628 have masses from $10^{3}$–$10^{4}$~M$_\odot$, with a wide range of extinctions peaking at $\sim 0.5$ and extending to values as high as $E(B{-}V) = 3$. Dashed histograms show the subset of eYSC sources identified as being embedded using the color selection criteria of \citet{linden23,linden24}. These embedded clusters are systematically younger and more reddened than the general eYSC population, confirming that this color selection provides an efficient diagnostic for identifying clusters in their early emerging phase, and suggests an evolutionary sequence for emerging clusters which clear their surrounding dust and gas as they evolve from eYSC-I to eYSC-II sources.}
\end{figure*}

\begin{figure*}[!htb]
\centerline{\includegraphics[width=\textwidth]{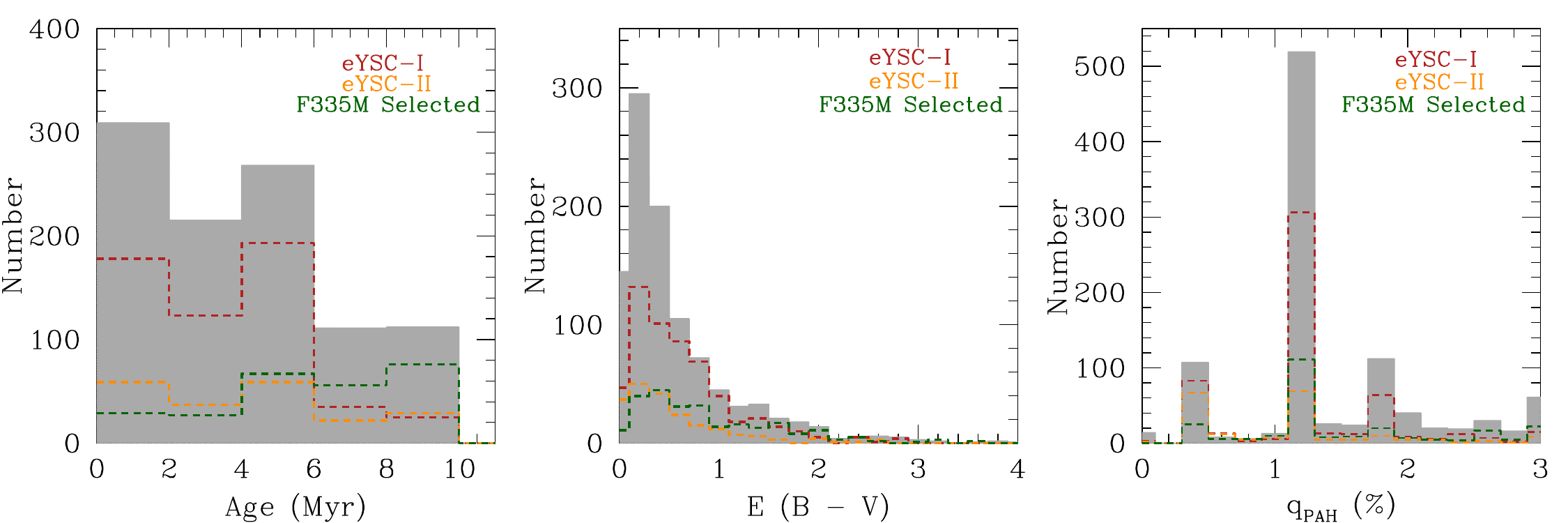}}
\caption{Comparison between the best-fit ages (\textbf{Left}), extinction $E(B{-}V)$ (\textbf{Middle}), and $q_{\mathrm PAH}$ (\textbf{Right}) of our F770W-selected sources (grey histogram) which overlap an eYSC-I, II, or F335M-selected source within 4 pixels. Using the same model grid adopted for our emerging cluster catalogs, the fits extend out to 7.7$\mu$m to include F560W and F770W observations with MIRI. In red, yellow, and green we show the individual distributions of eYSC-I, eYSC-II, and F335M-selected sources fit up to F444W from Figure~4. Overall we find that our MIRI-selected sources closely follow the distributions of age, extinction and $q_{\mathrm{PAH}}$ as we derive for eYSC-I sources using our NIRCam-only catalogs.}
\end{figure*}

\section{Results}

\subsection{Model Comparisons with the oYSC Catalog}

A comparison between the optically-selected cluster catalog produced from the LEGUS \citep{aa17} and FEAST surveys respectively is presented in Adamo et al. (in prep). Here we make a direct comparison between \textsc{cigale} and other stellar population models designed primarily to fit the SEDs of UV-optically bright sources. The \texttt{Yggdrasil} models have been widely adopted in studies of extragalactic star clusters, particularly within the LEGUS survey \citep[e.g.,][]{aa17,grasha17,mm18a,mm18b,bcw20,linden22a}, and therefore are used here for consistency with previous studies.

In the left Panel of Figure~2 we see that the median residuals of the \textsc{cigale} fits are minimal when restricting the wavelength coverage to $\lambda<2$\,$\mu$m. This is because \texttt{Yggdrasil} models do not include any contributions from hot dust or PAH emission, and are therefore unable to reproduce the observed mid-IR colors of our eYSC sample (see Figure~3). When both models are restricted to the same wavelength coverage we find that both approaches yield broadly consistent ages and masses (right Panels Figure~2). 

In pink we highlight the distribution of sources that have a best-fit extinction from \texttt{Yggdrasil} $\geq 0.2$ ($A_{\rm V} \sim 0.6$) as well as a best-fit metallicity from \textsc{cigale} of z=0.004. In the middle Panel of Figure~2 we see that the tail of sources with \texttt{Yggdrasil}-derived ages that are larger than the \textsc{cigale} fits is dominated by these sources with a lower-metallicity than the solar-metallicity tracks adopted in \texttt{Yggdrasil}. These discrepancies in the best-fit age are often referred to as ``catastrophic failures'' in SED fitting \citep{thilker25,whitmore23b}. Recent work has shown that such failures are common in young star cluster populations due to strong degeneracies between age, extinction, and metallicity, particularly when broad-band photometry is used \citep{thilker25,whitmore23b}. In such cases, $\sim 10-20\%$ of clusters in spiral galaxies can have inferred ages that are incorrect by an order of magnitude or more, where old globular clusters are frequently assigned intermediate ages of $\sim 100$ Myr or younger. This discrepancy in $E(B{-}V)$ also results in systematic offsets in cluster mass, seen in the tail of the pink distribution in the right Panel of Figure~2. The inclusion of additional JWST broad- and narrow-band filters as well as lower-metallically tracks can help break the degeneracies seen in previous studies.

Overall, these comparisons confirm that \textsc{cigale} provides a realistic framework for modeling the SEDs of young optically-selected cluster populations relative to existing SSP model frameworks.

\subsection{eYSC Physical Properties with \textsc{cigale}}

Figure~3 presents the F150W - F200W and F200W - F335M color--color distribution of the eYSCs in NGC~628 with the median values for eYSC-I, II, and F335M-selected sources shown as red, orange, and green squares respectively. We see that the bulk of the population lies significantly red-ward of the \texttt{Yggdrasil} evolutionary tracks, particularly in F200W - F335M, reflecting excess emission from hot dust and the 3.3\,$\mu$m PAH feature. In light blue we show a \textsc{cigale} evolutionary track made by convolving the model grid presented in Table 2 with the JWST NIRCam filters assuming an $E(B{-}V) = 0$. Here we see that \textsc{cigale} is able to broadly reproduce the observed colors of the youngest clusters when considering emission beyond $\lambda>2$\,$\mu$m, where purely stellar models like \texttt{Yggdrasil} clearly fail. The dashed region in Figure~3 marks the embedded-cluster selection (F200W - F335M $> 0$ and F150W - F200W $ > 0.47$) introduced by \citet{linden23,linden24}, which isolates clusters with the strongest PAH and nebular contributions. 

SED fitting of 2213 eYSC candidates in NGC~628 provides estimates of ages, stellar masses, and extinctions, as illustrated in Figure~4. The age distribution peaks strongly at $\sim$3--5~Myr, confirming that the catalog is dominated by very young sources. Masses are typically in the range $10^3$ - $10^{3.5}$~M$_\odot$, with a tail extending toward higher masses but a rapid drop-off beyond $10^4$~M$_\odot$. Extinctions are broadly distributed, with most clusters showing modest reddening of $E(B{-}V)\sim0.5$, but with a significant tail to $E(B{-}V)>2$ ($\sim$12\% of the sample). This is consistent with the results presented in \citet{graham25}, which demonstrate that deeply-embedded clusters represent a small fraction of the eYSC population in PHANGS galaxies. The dashed histograms in Figure~4 highlight the embedded cluster sub-samples selected by the NIR color criteria described above. The eYSC-I and II clusters identified as embedded sources are systematically younger ($\Delta_{\rm age} = 0.3-0.4$ dex), more reddened ($\Delta_{\rm E(B{-}V)} = 0.2$ dex), and more massive ($\Delta_{\rm age} = 0.1-0.2$ dex) than the general population of eYSCs. This result confirms that broad-band colors can provide an alternative means of isolating the earliest stages of cluster evolution when narrow-band observations are unavailable.

Overall, given the relatively uniform distribution of eYSC-I and II sources within the embedded cluster selection and $E(B{-}V)$ histogram shown in Figure 4, we infer that a strong excess in the F200W-F335M color is not directly linked to the reddening affecting the stellar continuum. One possible explanation is that clusters exhibiting red colors in both F150W-F200W and F200W-F335M host additional emission components, such as hot dust or enhanced nebular emission, which can significantly boost the flux in the F200W band. Alternatively, these trends may reflect the effects of a clumpy dust geometry in the surrounding interstellar medium \citep{whelan11}. In such a scenario, the measured $E(B{-}V)$ toward the stellar cluster depends sensitively on the specific line of sight through a non-uniform dust distribution. In particular, systems with high dust covering fractions may exhibit systematically low apparent extinction if the observed emission is dominated by less extincted lines of sight. Overall, the observed emission represents an average over multiple dust structures within our photometric aperture. This naturally leads to a mixing of eYSC-I and eYSC-II populations in near-IR color space, producing the observed overlap in their NIR properties despite differences in their evolutionary state.

\subsection{F770W-Selected Sources with \textsc{cigale}}

The role of mid-infrared 7.7$\mu$m PAH emission in tracing embedded stars clusters is examined in Figure~5. Here we compare the physical properties of F770W-selected clusters with those of eYSC-I, eYSC-II, and F335M-selected clusters that overlap an F770W source within four pixels. The gray histograms are the resulting distributions with F560W and F770W photometry from MIRI included in the fit with \textsc{cigale}. The dashed red, orange, and green histograms show the same overlapping sources fit with our NIRCam only catalog (see Figure~4). Overall, we find that the F770W-selected catalog closely follow the age distribution of eYSC-I sources. 

In the middle Panel of Figure~7 we see that $\sim$4\% of F770W-selected clusters  show $E(B{-}V)>2$, in agreement with the distributions of eYSC-I, II, and F335M-selected sources fit out to F444W. Importantly, we find that only $\sim$2\% of sources with ages $<4$~Myr ($\sim$10 objects in total) lack a counterpart in eYSC-I, eYSC-II, or F335M within 4 pixels. In other words, there is no evidence for a large hidden population of clusters that are bright at F770W but entirely missed by our shorter-wavelength observations.

In the right Panel of Figure~7 we see that the distribution of $q_{\mathrm PAH}$ for sources including both the 3.3$\mu$m (F335M) and 7.7$\mu$m (F770W) PAH features closely follows the distribution of eYSC-I and II sources. This result demonstrates that fits which only include F335M observations of  3.3$\mu$m PAH emission do not systematically bias the results from \textsc{cigale} SED fitting.

\section{Discussion}

\subsection{Cluster Emergence Timescales}

Our results suggest that eYSCs in NGC~628 represent a transitional stage between fully embedded stellar populations and optically visible clusters. Their near-IR excess and association with PAH emission are consistent with feedback-driven clearing of their birth material on short ($\lesssim 5-7$~Myr) timescales. The increasing fraction of embedded clusters toward younger ages (left panel of Figure~4), together with the strong PAH emission observed in clusters younger than 3 Myr, suggests that once massive stars form, stellar feedback rapidly heats and destroys small dust grains.

To quantify these emergence timescales, we compared the relative numbers of emerging (eYSC) and optically-visible (oYSC) clusters and calculate two characteristic values: 
\[
t_{tot} = 10 \times \frac{N_{\mathrm{eYSC I + II}}}{N_{\mathrm{eYSC I + II}} + N_{\mathrm{oYSC}}(<10~\mathrm{Myr})},
\]
and
\[
t_{PDR} = 10 \times \frac{N_{\mathrm{eYSC I}}}{N_{\mathrm{eYSC I + II}} + N_{\mathrm{oYSC}}(<10~\mathrm{Myr})}.
\]

Where $t_{tot}$ represents the timescale for the full cluster population to emerge and become optically visible, and $t_{PDR}$ is the timescale for clusters to transition from being fully-embedded to partially-embedded as they continue to emerge. For the eYSC catalogs our default approach is to take the total numbers of eYSC-I and II sources with fitted $\chi_{\rm red}^{2} < 20$ (1335 sources). For the optical-selected catalog we take all oYSC with fitted ages $\leq 10$ Myr, $\chi_{\rm red}^{2} < 20$, and a requirement to fall within the JWST footprint (1329 sources). Finally, we find that a  few percent of the
eYSCs overlap in position an optical oYSC source (within 4 NIRCam pixels). For this analysis we removed these objects from the eYSC catalogs and add them to the oYSC catalogs resulting in 1129 and 1535 sources respectively.

Using these cuts we find $t_{tot} \approx 5$~Myr and $t_{PDR} \approx 3$~Myr. Repeating these calculations with catalogs fit only to $\lambda<2\,\mu$m yields identical results ($t_{tot} \approx 5$~Myr and $t_{PDR} \approx 3$~Myr), demonstrating that the derived timescales are robust to our choice of wavelength coverage when fitting the SEDs. 

Alternative sample selections, including removing the $\chi_{\rm red}^{2}$ cut on eYSCs or oYSCs, yield somewhat longer timescales, with $t_{tot} \approx 6$~Myr and $t_{PDR} \approx 4$~Myr for the full sample. Finally, to mitigate issues with observational completeness, we apply a mass cut of $M \geq 10^{3} M_{\odot}$ to both the eYSC and oYSC catalogs. This additional cut yields timescales for $t_{tot}$ and $t_{PDR}$ of 7 and 4 Myr respectively, which is consistent with the results for the full sample of eYSC and oYSCs in the FEAST survey \citep{pedrini26}. In all cases, the same qualitative picture emerges: the embedded phase is short-lived (4~Myr for the bulk of clusters), and by $\sim 7$ Myr the population of eYSCs has become optically-visible. 

These results are also broadly consistent with recent STARFORGE simulations of cluster emergence \citep{wainer25}, which find that the transition from embedded to exposed is fast for individual massive stars ($\sim 2$ Myr). Once these massive stars are revealed, their localized, pre-supernova feedback then impacts the cloud, driving gas clearance. As such, dust emission appears to dominate the luminosity for $\sim 2$ Myr before the H$\alpha$ luminosity becomes the brightest component of the cluster SED. Finally, because the initial embedding of the most luminous stars in clusters is highly localized, the emergence of massive stars is faster than the overall disruption of the cloud by gas expulsion. Both simulations and observations suggest that the embedded cluster phase, where dust emission can be used to uniquely identify sources, is short-lived in nearby galaxies.

An additional consistency check on these inferred emergence timescales can be obtained by comparing the implied cluster formation rates in each evolutionary phase with the total star formation rate (SFR) of the galaxy. In a steady-state scenario where clusters evolve through a continuous sequence from eYSC-I to eYSC-II and finally to optically visible oYSCs, the total stellar mass formed in each phase divided by the corresponding phase lifetime should yield the same formation rate. However, if a significant fraction of clusters are disrupted or dispersed during their early evolution, these relations will systematically underestimate the true durations of the embedded phases. In this case, correction factors accounting for cluster mass loss would need to be incorporated into the number counts used in Equations (1) and (2). Cluster disruption rates ($dN/dt \propto t^{- \beta}$) have been measured in a variety of nearby galaxies with HST finding values of $\beta$ that range from 0.1-1 \citep{aab15}. Taking 1 as our upper-limit for $\beta$ implies a loss of 90\% of the cluster population for each successive dex in age, and that the survival fraction scales as $1/t$. As a result, the correction to recover the true number of clusters depends on the age separation between the evolutionary phases being compared. The median difference in age between eYSC-I and eYSC-II is $\sim 1$ Myr, and therefore $t_{\rm PDR}$ should be robust to assumptions in the cluster disruption rate. A full assessment of the effects of cluster disruption on the resulting population of eYSC and oYSCs is left for future work.

\begin{figure*}[!htb]
\centerline{\includegraphics[width=\textwidth]{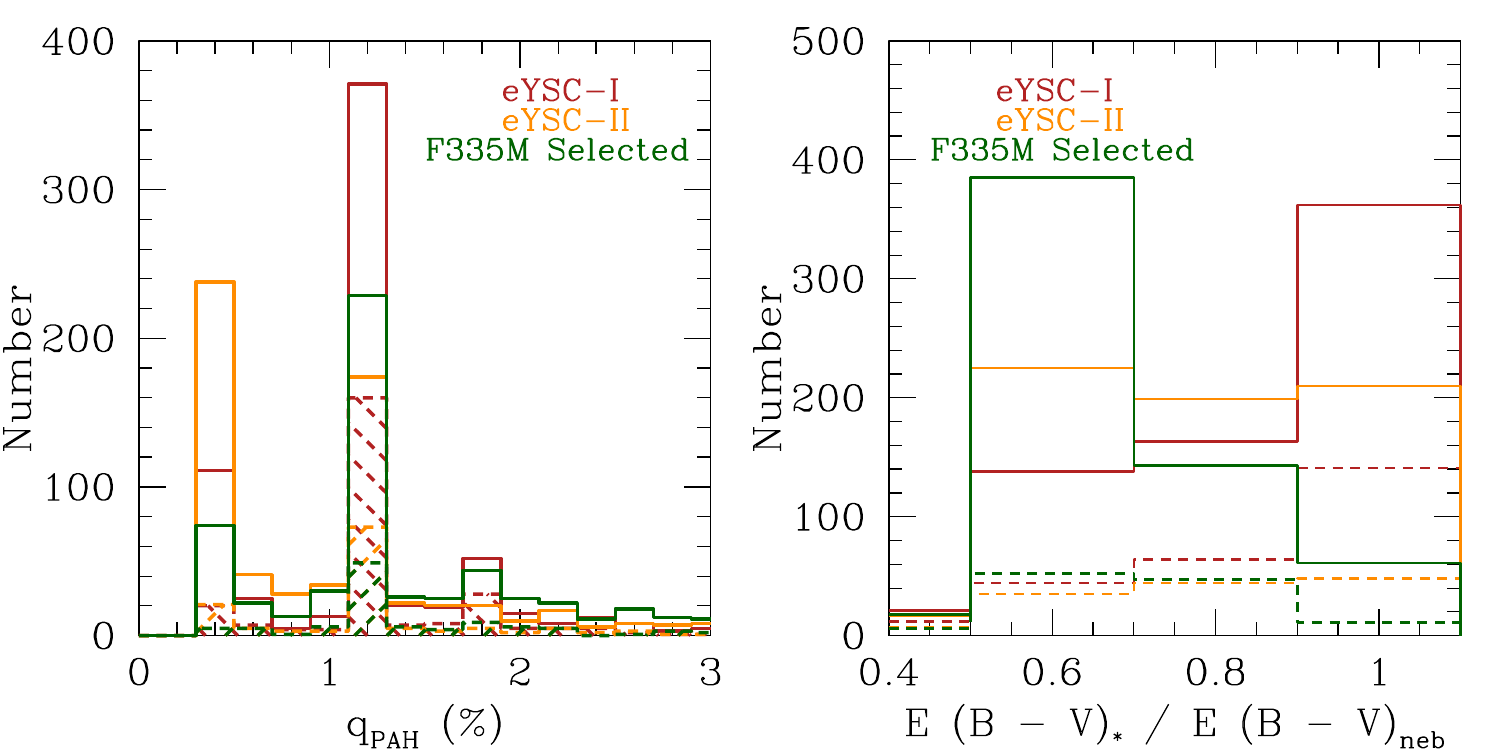}}
\caption{Distributions of dust and attenuation properties for emerging clusters in NGC~628 derived from \textsc{cigale}. \textbf{Left:} Fractional PAH abundance ($q_{\mathrm{PAH}}$) for clusters identified as eYSC-I (red), eYSC-II (orange), and F335M-selected (green) sources. The distribution of $q_{\mathrm{PAH}}$ peaks at systematically lower values relative to eYSC-I sources. \textbf{Right:} Ratio of nebular to stellar continuum attenuation, $E(B{-}V)_{\mathrm{\star}}/E(B{-}V)_{neb}$, which traces the relative obscuration of ionized gas compared to stars. Dashed histograms in both Panels show the subset of embedded clusters identified via the color selection of \citet{linden23,linden24}. These embedded clusters preferentially exhibit elevated PAH fractions and higher nebular-to-stellar attenuation ratios, reinforcing the interpretation that they remain partially-enshrouded in their dense birth material before evolving along the emerging sequence we propose for eYSC-I and eYSC-II sources.}
\end{figure*}

\begin{figure*}[]
\centerline{\includegraphics[]{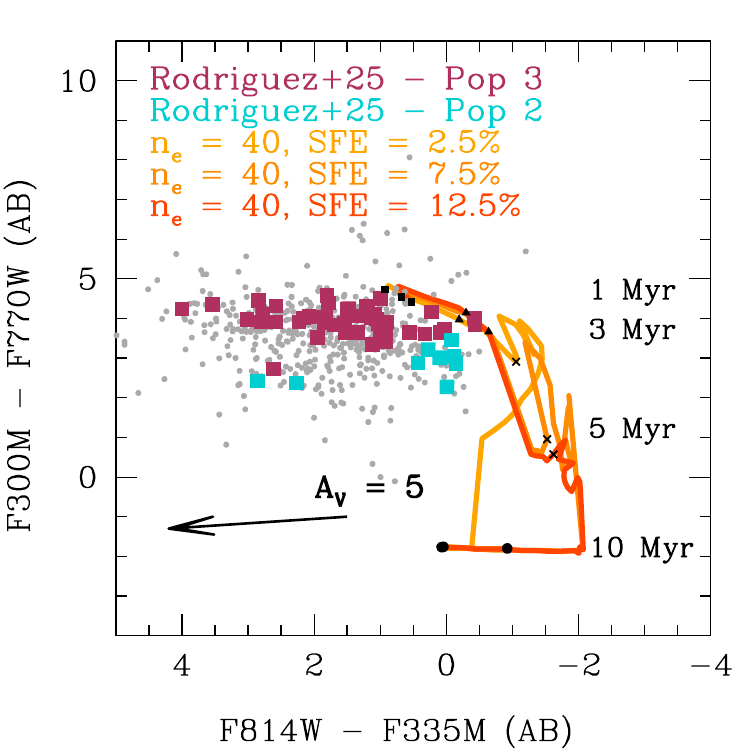}}
\caption{Near- and mid-IR color--color comparison of our F770W-selected sources which overlap eYSC-I, II and F335M-selected sources within 4 NIRcam pixels (gray points). Magenta and turquoise squares are $3.3\mu$m-bright sources identified \citet{rodriguez25}. The distribution of F770W-selected sources closely aligns with the compact 3.3\,$\mu$m PAH emitters identified by \citet{rodriguez25}, confirming that F770W can be used to isolate similarly young and dust-reddened clusters. For comparison, evolutionary tracks from the \texttt{TODDLERS} SED models with $E(B{-}V) = 0$, a fixed electron density ($n_{e} = 40$), and star formation efficiencies that range from 2-12\% are shown \citep{kapoor23}. Locations of 1, 3, 5, and 10 Myr are given as square, triangle, cross, and circular symbols for each model respectively. We find good agreement between the observed locus of young clusters ($\lesssim 5$~Myr) and the \texttt{TODDLERS}  model colors, including the rapid fading of 3.3$\mu$m PAH emission at later times.}
\end{figure*}

\subsection{Optical and Near-Infrared Dust Properties}

Dust properties derived from \textsc{cigale} fits provide additional constraints on cluster environments (Figure~6). In the left Panel we see that eYSC-II sources have on average lower $q_{\mathrm{PAH}}$ values than eYSC-I sources. Further, we see that embedded clusters identified via the \citet{linden23,linden24} selection (dashed histograms) tend to have elevated $q_{\mathrm{PAH}}$ relative to less-embedded counterparts in all three cluster catalogs. \citet{egorov25} find that the PAH fraction ($R_{\rm PAH}$) toward H\textsc{ii} regions is strongly anti-correlated with the local ionization parameter, suggesting that PAH destruction is correlated with hydrogen-ionizing UV radiation. Although our results appear to be consistent with this picture, longer-wavelengths observations of PAH features which probe larger dust grains are required, to verify this result for individual eYSCs.

Overall, the distribution of fractional PAH abundances for eYSC-I, eYSC-II, and F335M-selected clusters generally exhibit lower values ($q_{\mathrm{PAH}} < 2 \%$) compared to typical kpc-scale star-forming regions in nearby galaxies \citep[$q_{\mathrm{PAH}} \sim 3-6 \%$][]{sutter24}. These results are broadly consistent with a recent study of 24,945 compact mid-IR peaks in 19 nearby galaxies from PHANGS which demonstrated that ISM sources, defined by the presence of PAH emission detected in both the NIRCam and MIRI bands, have observed $q_{\mathrm{PAH}}$ values between 1\% and 2\% \citep{hassani25}.

The right panel of Figure~6 presents the ratio of stellar-to-nebular attenuation, $E(B{-}V)_{\mathrm{\star}}/E(B{-}V)_{neb}$. While the full cluster sample spans a broad range of values, the embedded eYSC-I and II shows systematically higher ratios, indicating that ionized gas emission lines remain more strongly obscured relative to the stellar continuum. These results indicate that embedded clusters appear partially enshrouded by their natal gas, where dust geometry and clumpiness preferentially obscure ionized regions, and that as clusters evolve from being embedded to dust-free, or from eYSC-I to eYSC-II sources the stellar-to-nebular attenuation ratio decreases \citep{indebetouw06,whelan11}. This result is also consistent with \citet{scheuermann23} which used observations of 24,000 H\textsc{ii} regions across 19 galaxies in PHANGS to demonstrate that the ratio $E(B{-}V)_{\mathrm{\star}}/E(B{-}V)_{neb}$ decreases from $\sim 1$ to 0.3 as the underlying stellar population ages increase from 1 - 6 Myr.

Taken together, the elevated $q_{\mathrm{PAH}}$ values and higher $E(B{-}V)_{\mathrm{\star}}/E(B{-}V)_{neb}$ ratios, for sources selected to be embedded suggests that the combination of strong PAH emission and high relative nebular obscuration provides a coherent picture of clusters in transition: massive stars have already formed and begun to ionize their surroundings, which rapidly destroys the surrounding PAH-emitting grains or clears from our photometric apertures, as clusters evolve from eYSC-I to eYSC-II while feedback disperses the natal material.

\subsection{Comparisons between NIRCam- and MIRI-Selected Sources}

Bright 3.3\,$\mu$m PAH emitters identified at $\sim$10~pc resolution have been shown to trace extremely young, dust-enshrouded cluster candidates that are often under-represented in optical catalogs \citep{rodriguez25}. \citet{rodriguez25} identify 63 sources (labeled "Pop 3" and "Pop 2") as having strong ($\geq 5 \sigma$) 3.3 $\mu$m PAH emission. In Figure~7 we directly compare the two samples using F300M-F770W (a tracer of strong 7.7$\mu$m PAH emission) and F814W-F335M (a tracer of strong 3.3$\mu$m emission) colors. We see that the F770W clusters that overlap with eYSC-I, eYSC-II, or F335M sources occupy the same locus as 3.3\,$\mu$m PAH emitters identified by \citet{rodriguez25}, confirming that both selections independently trace very young, dusty clusters on $\sim$10~pc scales.

Further, in Figure~7 we over-plot models from the Time evolution of Observables including Dust Diagnostics and Line Emission from Regions containing young Stars (\texttt{TODDLERS}) library \citep{kapoor23}. Unlike \textsc{cigale}, \texttt{TODDLERS} explicitly models different stellar feedback processes including stellar winds, supernovae, and radiation pressure, as well as the gravitational forces on the gas. Time-dependent UV–mm SEDs with varying metallicity, star-formation efficiency (SFE), and birth cloud density ($n_{e}$) are then constructed with the  radiative transfer code SKIRT. We see that regardless of the SFE adopted for the birth-cloud \texttt{TODDLERS} models show good agreement with the observed colors of clusters younger than 5~Myr, and reproduces the rapid fading of PAH emission at later times. This agreement strengthens the conclusion that the strongest mid-IR emission is confined to the earliest embedded phases, with PAHs dispersing quickly as feedback clears natal material \citep{kapoor23}. Overall, we find that F770W can also be used as an efficient and independent method to identify eYSC populations in nearby galaxies.

\begin{figure*}[!htb]
\centerline{\includegraphics[width=\textwidth]{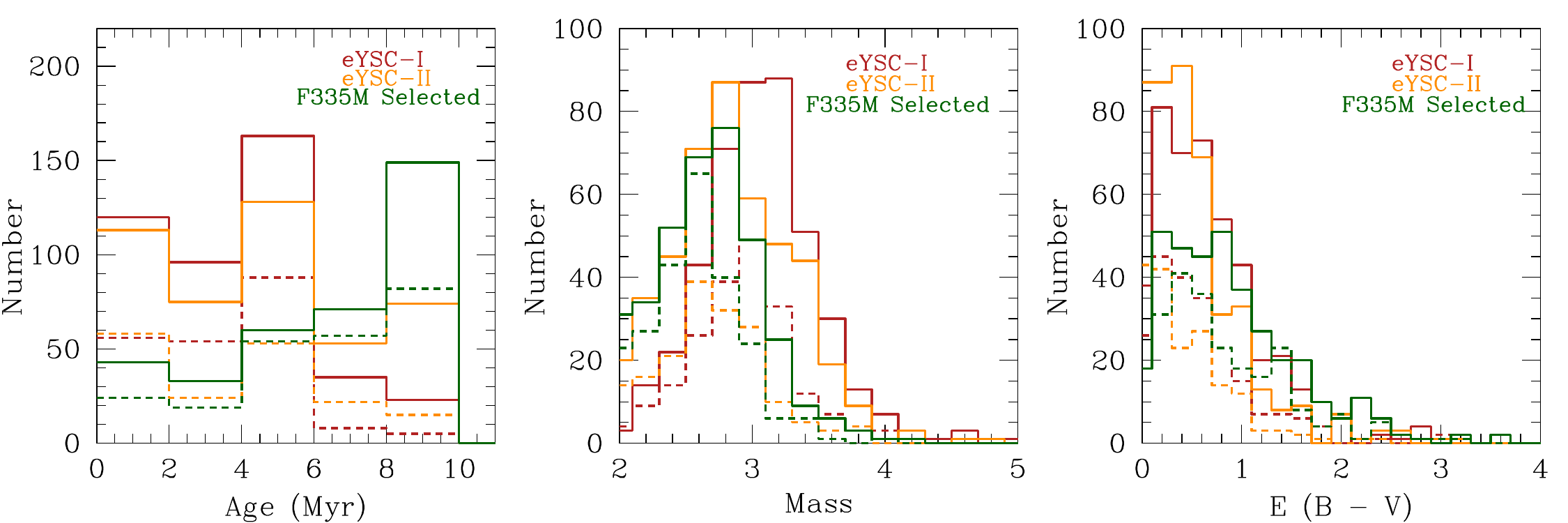}}
\caption{Comparison of the derived eYSC best-fit ages (\textbf{Left}), stellar masses (\textbf{Middle}), and extinction $E(B{-}V)$ (\textbf{Right}) for sources falling in arm (solid lines) and inter-arm (dashed-lines) regions using the environmental mask of \citet{querejeta21}. Although the age distributions appear broadly consistent, star clusters that reside in the spiral arms of NGC~628 appear to be systematically more massive ($\sim$0.1--0.2 dex) and more dust extincted ($\sim$0.1--0.2 dex) relative to inter-arm clusters. These results suggest that spiral arm regions preferentially host more massive and more embedded young clusters.}
\end{figure*}

\subsection{Environmental Dependence}

Using the environmental mask presented in \citet{querejeta21} for NGC 628 we can divide our eYSC and oYSC catalogs into sources which reside in either arm or inter-arm environments (Figure~8). When examining the median age of each eYSC catalog we see only modest variations: eYSC-I sources have nearly identical ages in the arms and inter-arms ($\sim$4.0~Myr), while eYSC-II clusters are slightly older in the arms (4.9~Myr) than in the inter-arm regions (4.1~Myr), and F335M-selected sources have median ages of $\sim$6.0~Myr in both environments (left Panel Figure~8). The emergence timescales for arm and inter-arm regions suggest that cluster emergence proceeds at comparable rates across environments in NGC 628, although perhaps slightly faster in the lower density environments of the inter-arms. For arm regions, we find $t_{tot} \approx 3.4$--6.0~Myr and $t_{PDR} \approx 1.7$--3.1~Myr, while in inter-arm regions the corresponding timescales are slightly shorter: $t_{tot} \approx 2.3$--4.3~Myr and $t_{PDR} \approx 1.3$--2.5~Myr. This result is broadly consistent with the result of \citet{buckner26}, which quantify the spatial clustering of eYSC and oYSCs in NGC 628 finding that isolated eYSCs in lower gas density environments disperse faster.

Further, when examining the median cluster mass of each distribution, we find for eYSC-I, the median mass in the arms is $\log(M/M_{\odot}) \approx 3.10$ compared to 2.97 in the inter-arm regions (middle Panel Figure~8). Similarly, eYSC-II clusters have median masses of $\log(M/M_{\odot}) = 3.03$ in the arms and 2.86 in the inter-arm regions, and F335M-selected sources have a median mass of 2.80 and 2.68 respectively. In all cases, spiral arm clusters are $\sim$0.1--0.2 dex more massive on average, consistent with the higher-pressure environments of the arms favoring the formation of more massive clusters \citep{dobbs11}. Finally, when examining the median $E(B{-}V)$ for eYSC-I eYSC-II, and F335M-selected sources in NGC 628 we find that clusters in arms are $\sim$0.1--0.2 dex more extincted on average (right Panel Figure~8).

These comparisons demonstrate that while cluster ages and cluster emergence timescales are broadly similar between environments, clusters in spiral arms tend to be more massive, more dust reddened, and take longer to emerge. FEAST observations of M83 demonstrate that when galaxies display more prominent spiral arms and central bars, these differences are further magnified \citep{knutas25}.

\subsection{Comparisons with Recent Cluster Studies}

Empirical SED templates built from HST and JWST observations indicate that emerging star clusters older than $\sim$5~Myr no longer display strong PAH or infrared dust excesses \citep{whitmore25,henny25}. In NGC~628 we find that eYSC-I sources have best-fit SEDs which include contributions from hot dust and PAH emission, and our embedded-phase clearing estimates (2--5~Myr) are consistent with the notion that strong PAH emission is a short-lived phase in the evolution of compact clusters. This supports a scenario where radiative and mechanical feedback efficiently destroy and/or process small grains within a few Myr of formation.

With NIRCam and MIRI observations of the $3.3\mu$m, $7.7\mu$m, and $11.3\mu$m PAH features for star-forming regions across 19 systems, \citet{dale25} finds that the observed PAH ratios indicate highly ionized and larger grain populations reside near optical stellar clusters, and that the 3.3\,$\mu$m feature is suppressed in regions exposed to harder radiation fields. FEAST observations of ionized gas and PAH emission also reveal that PAH ratios in proximity to eYSC populations are further suppressed relative to optical clusters \citep{pedrini24,gregg24}.


\subsection{Limitations and Future Work}

While our analysis highlights the strength of combining HST and JWST imaging with \textsc{cigale} modeling to characterize emerging cluster populations, several limitations remain. First, \textsc{cigale} assumes deterministic sampling of the stellar IMF. This is appropriate for sufficiently massive clusters ($M \gtrsim 10^{4}$~M$_{\odot}$), where the IMF can be considered fully populated, but it becomes less valid at lower masses ($M < 10^{4}$~M$_{\odot}$), where stochastic IMF sampling become important. Such clusters may fail to form high-mass stars, effectively truncating the IMF, and resulting in systematically fainter sources with a larger scatter compared to a fully sampled population \citep{barbaro77,girardi93,cervino06,fumagalli11,fouesneau12,hannon19}. In color--color space, these clusters may appear bluer in $V{-}I$ than the predictions of standard SSP tracks at ages of 1--5~Myr due to a lack of red post-main sequence stars. Conversely, when an excess of evolved stars is present, clusters can appear redder than the SSP tracks. The latter effect is degenerate with dust reddening, and therefore can bias SED fits toward higher inferred $E(B{-}V)$ values \citep{fouesneau12,hannon19}. Incorporating probabilistic IMF sampling into SED fitting frameworks \citep[e.g, SLUG -][]{krumholz15a} will be crucial for future studies that probe the faint end of the cluster mass function (Pedrini et al. in prep). 

Further, the grain size distributions and selective destruction of smaller PAHs are not modeled in \textsc{cigale}. The PAH emission is treated as a fractional abundance, $q_{\mathrm PAH}$), where the relative PAH size distribution does not change with age or local radiation field hardness. Additionally, the assumption of energy balance (i.e. that absorbed starlight equals re‐emitted dust emission) that \textsc{cigale} adopts becomes less valid when gas and dust are being removed or destroyed on short timescales. \citet{pedrini25} demonstrate that for clusters older than $\sim 5$ Myr, \textsc{cigale} tends to over-predict the PAH and hot dust emission, likely as a result of this requirement. Overall, while our current models capture a broad range of cluster properties, these missing components may introduce systematic uncertainties that are not yet able to be captured in \textsc{cigale}.

\section{Summary}

In this work we have utilized observations from HST and JWST to identify over 12,000 optically-selected YSCs, eYSCs, and MIRI-selected candidates in NGC~628. We present our choices for the \textsc{cigale} modeling framework adopted to fit the combined SEDs and derive star cluster physical properties in the FEAST survey. Our main findings are:

\begin{enumerate}[leftmargin=*]
    \item Emerging star clusters, selected based on a visual inspection of Pa$\alpha$, Br$\alpha$, and 3.3$\mu$m PAH emission maps have an age distribution that peaks at 3--5~Myr. Only a small fraction ($\sim$6\%) are older than 6~Myr.
    \item The majority of detected eYSCs in NGC 628 are moderately massive ($10^{3}$--$10^{4}$\,M$_{\odot}$), with a steep decline toward higher masses.
    \item Roughly 12\% of eYSCs are highly reddened with $E(B{-}V)>2$, indicating that they are still significantly embedded.
    \item Near-infrared photometry is essential to constrain hot dust and PAH emission in eYSCs. These components dominate the SEDs of clusters younger than 5~Myr, and are strongly correlated with the observed F200W-F335M colors of eYSC-I and II sources.
    \item Clearing timescales estimated from the ratio of emerging to optically-selected young star clusters ($t < 10$Myr) with masses $M > 10^{3} M_{\odot}$ are found to be short ($t_{tot} \approx 7$~Myr and $t_{PDR} \approx 4$~Myr), with a weak dependence on residing in arm or inter-arm regions. However, star clusters identified in spiral arms are on average more massive, more reddened, and have more dust emission than inter-arm clusters.
    \item Cross-comparison with MIRI F770W-selected sources reveals that 65\% of eYSC-I, 27\% of eYSC-II, and 40\% of F335M-selected clusters overlap a MIRI peak within 4 pixels, with a steep decline at ages $>5$~Myr. This result confirms that many eYSCs coincide with strong mid-IR emission.
    \vspace{0.1in}
\end{enumerate}

These results demonstrate the power of combining HST and JWST imaging in the FEAST survey to study the earliest phases of cluster evolution. Further, we have shown that while a number of improvements can still be made, \textsc{cigale} provides the necessary framework to interpret the SEDs of eYSCs in nearby galaxies. Future work combining these datasets with spectroscopy will provide further insights into the interplay between stellar populations, ionized gas, and dust during the critical emergence phase of cluster evolution.

\begin{acknowledgements}

This work is based on observations made with the NASA/ESA/CSA James Webb Space Telescope, which is operated by the Association of Universities for Research in Astronomy, Inc., under NASA contract NAS 5-03127. The Feedback in Emerging extrAgalactic Star clusTers data (GO1783) presented in this article were obtained from the Mikulski Archive for Space Telescopes (MAST) at the Space Telescope Science Institute. The specific observations analyzed can be accessed via \dataset[doi:10.17909/6dc1-9h53]{https://doi.org/10.17909/6dc1-9h53}. HFV and AA acknowledge support from the Swedish National Space Agency (SNSA) through the grant 2023-00260. AA and AP acknowledge support from SNSA through grant 2021-00108. ADC acknowledges the support from a Royal Society University Research Fellowship (URF/R1/19160 and URF/R/241028). KG is supported by the Australian Research Council through the Discovery Early Career Researcher Award (DECRA) Fellowship (project number DE220100766) funded by the Australian Government. 

\end{acknowledgements}

\bibliography{master_ref}{}

\appendix

\subsection{Comparison with Fits out to 2$\mu$m}

A key test of our SED fitting framework is to assess the extent to which JWST long-wavelength coverage improves the accuracy and robustness of the derived cluster parameters. To evaluate this, we repeated the \textsc{cigale} fits of our eYSC catalog using only photometry for $\lambda \leq 2\mu$m, corresponding to the combination of HST filters and JWST NIRCam short-wavelength bands. The resulting cluster ages, masses, and extinctions are shown in Figure~9, and are compared to the full 0.3--4.4$\mu$m fits in Figure~2.

Overall, we find that when dust emission models are excluded and the fits are restricted to $\lambda \leq 2\mu$m, the resulting age and extinction distributions closely resemble those obtained from the full \textsc{cigale} fits (left and right Panels of Figure~9). However, we also find that the inferred stellar masses are systematically higher ($\sim 0.4$ dex) than seen in the middle Panel of Figure~4, and exhibit larger scatter. This is likely a consequence of the energy balance assumption: in the absence of an explicit dust emission component, the flux at longer wavelengths must be attributed entirely to stellar emission. Because the age range is restricted to $\leq$10 Myr, the models cannot reproduce these colors by shifting to older populations, and instead compensate by increasing the stellar mass normalization.

Further, when the dust emission models are included but the fit is restricted to $\lambda \leq 2\mu$m, the resulting age distributions shift toward younger values ($\sim$1--2 Myr). This behavior likely arises because the dust component is poorly constrained in the absence of longer-wavelength data. In particular, the models can reproduce red near-IR colors (e.g., in F200W) by favoring solutions with high $q_{\mathrm PAH}$ values, effectively compensating for the lack of mid-infrared constraints. As a result, the fitting procedure can artificially select very young, dust-rich best-fit solutions. This highlights a key degeneracy between stellar reddening and dust emission in the near-IR, and underscores the importance of including PAH-sensitive bands (e.g., F335M and longer wavelengths) to robustly constrain dust properties when using the dust emission models in \textsc{cigale}.

Together, these tests demonstrate that the optical and NIR-only fits can recover the broad stellar population properties of eYSCs. However, important systematic effects emerge when considering the role of dust emission for breaking degeneracies between age and extinction, and characterizing the youngest and most dust-embedded cluster populations.

\begin{figure*}[!htb]
\centerline{\includegraphics[width=\textwidth]{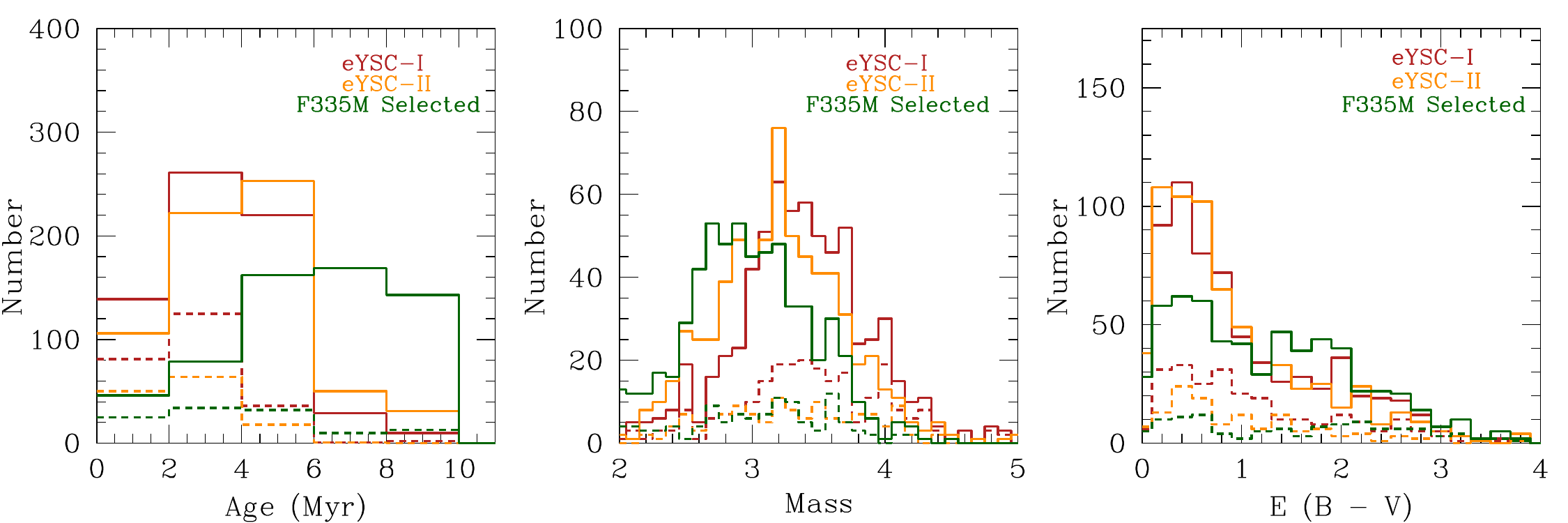}}
\caption{Comparison of the derived eYSC best-fit ages (\textbf{Left}), stellar masses (\textbf{Middle}), and extinction $E(B{-}V)$ (\textbf{Right}) using SED fits restricted to $\lambda < 2$\,$\mu$m. The distribution of ages for eYSC-I and II sources peaks at $\sim 2$ Myr, which is slightly younger than the median derived for the full 0.3-4.4$\mu$m fits. We further see similar distributions for the derived cluster masses and dust extinctions, and when isolating embedded eYSC sources we see that the median age, mass, and extinction distributions agree between our full \textsc{cigale} SED fitting and our fits restricted to $\lambda < 2$\,$\mu$m.}
\end{figure*}


\subsection{Residuals of \textsc{cigale} Fitting including MIRI}

To further evaluate the robustness of our model-fitting approach, we examined the residuals of our standard \textsc{cigale} fits from $0.3-4.4\mu$m (left Panel Figure~10) relative to the residuals when MIRI F560W and F770W photometry is included (right Panel Figure~10). The mid-IR data provides a more stringent test of whether the adopted dust models can reproduce the observed emission at 3--8\,$\mu$m, where PAH features dominate.

We find that including the MIRI bands slightly reduces the overall scatter from 3-8 $\mu$m for eYSC-I, eYSC-II, and F335M-selected sources, with typical residuals of $\lesssim 0.2$~dex. In particular, the scatter in the broad- (F200W and F444W) and medium-band (F300M and F360M) continuum filters for F335M-selected sources is significantly reduced. Without MIRI data, the models tend to under-predict the flux in these bands (positive fractional residual). However, when examining the F335M and F770W residuals when MIRI photometry is included we see that the overall scatter in F335M has increased with models tending to under-predict (positive residuals) the 3.3$\mu$m PAH feature while simultaneously over-predicting (negative residuals) the 7.7$\mu$m PAH feature. Despite this, the overall $q_{\mathrm{PAH}}$ distribution for NIRCam-only and MIRI-selected sources appear to be the same (right Panel Figure~5).

These deviations between the models and observations may represent clusters with unusually high PAH abundances or more complex geometries than captured by our simple prescriptions. Three-dimensional radiative transfer models have demonstrated that changing both the degree of clumpiness and the viewing angle can result in substantial variations in the near- and mid-IR SEDs of embedded clusters \citep{whelan11}. The observed excess emission in some systems may therefore arise from line-of-sight effects and unresolved clumpy structure, rather than requiring intrinsically anomalous dust properties. Future work combining the photometry with mid-IR spectroscopy will be necessary to disentangle these effects that are not fully captured in our more simplified modeling framework.

\begin{figure*}[!htb]
\centerline{\includegraphics[width=\textwidth]{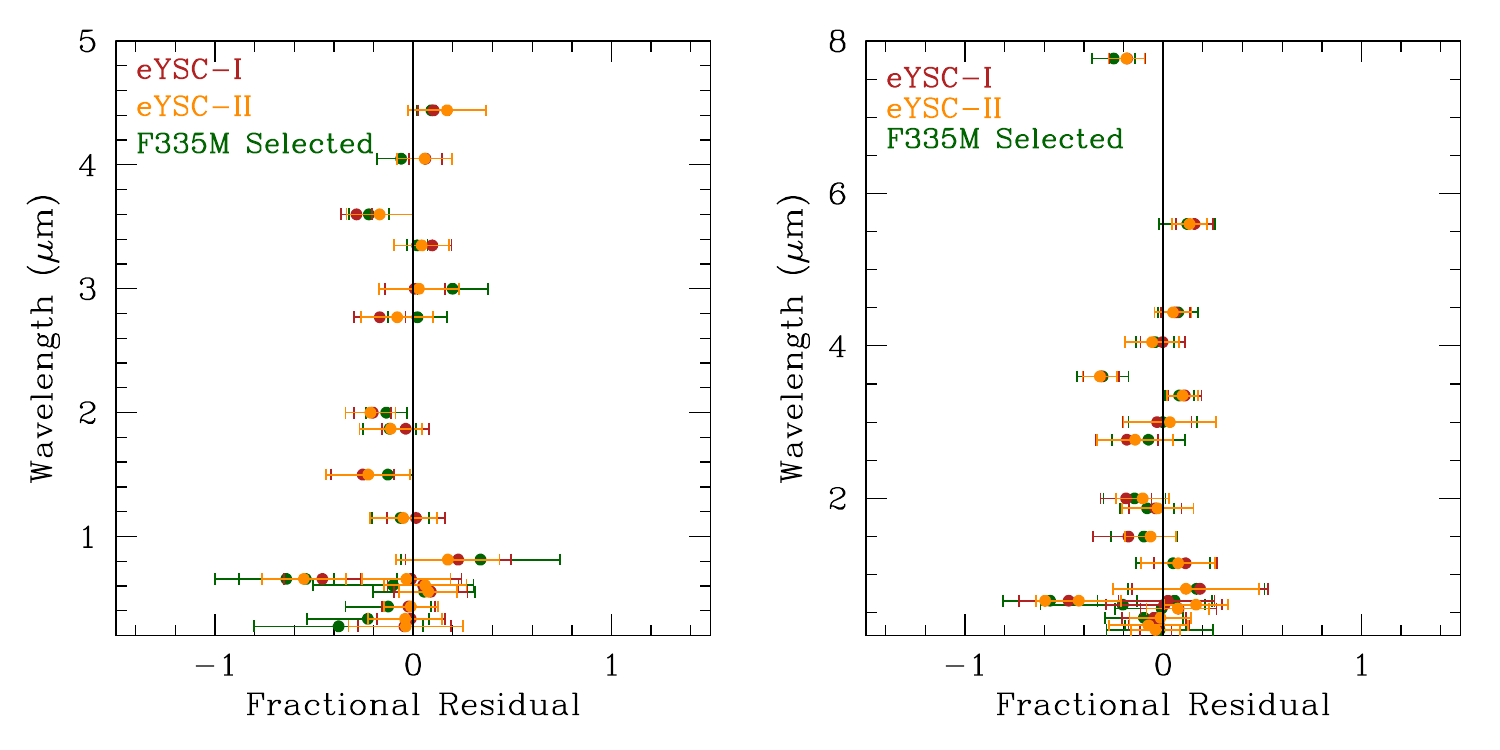}}
\caption{Residuals from SED fitting when including all HST and JWST NIRCam observations (\textbf{Left}), as well as JWST MIRI photometry (\textbf{Right}). The addition of mid-IR F560W and F770W observations help to reduce the overall scatter from 3--8\,$\mu$m, and confirms that hot dust and PAH emission must be included to reproduce the youngest clusters.}
\end{figure*}

\end{document}

%% file: catalog_numbers.tex
\begin{table}
\centerline{Detection Statistics and Final Catalogs}
\centering
\begin{tabular}{lll}
\hline
\hline
Catalog        & Total Detections & Final Catalog \\
\hline
eYSC-I         & 835 & 684 \\
eYSC-II         & 741 & 651 \\
F335M-selected           & 637 & 607 \\
F770W-selected           & 1328 & 1023 \\
Optically-selected         & 9197 & 1329 \\                        
\hline \hline
\end{tabular}
\caption{Total eYSC detections represents the number of sources identified with Sextractor, and the final catalogs are the total number of sources that have a CIGALE-fitted $\chi^{2} \leq 20$. For the final MIRI-selected catalog we apply the same $\chi^{2} \leq 20$ cut, and require sources to match a source in either the eYSCI, II, or F335M-selected catalogs within 4 MIRI pixels. For the final optically-selected catalog we apply the same $\chi^{2}$ cut, and additionally require all sources to be detected within our JWST NIRCam footprint and have machine-learning based optical classifications of 1, 2, or 3.}
\end{table}

%% file: table_parameters_class123.tex
\begin{table*}
\centerline{SED Model Parameters Emerging Star Clusters}
\centering
\begin{tabular}{ll}
\hline
\hline
Star Formation History          & Instantaneous Burst\\
Reddening \& Extinction         & $E(B{-}V)=$~[0 - 5]~mag; $\Delta=0.1$~mag; $R_V=A_V/E(B{-}V)=3.1$\\
Age                            & 1, 2, 3, 4, 5, 6, 7, 8, 9, 10 \\ 
SSP model                       & Bruzual \& Charlot 2003 \\
Metallicity                     & $Z=0.02$ \\
IMF                             & Chabrier et al. 2003; [1 - 100]~$M_\odot$; fully sampled\\
$n_{\rm e}$                                 &10, 100 \\
Log(U)                                  &-3.5, -3.0, -2.5, -2.0\\
$f_{\rm esc}$                                 &0.01, 0.1, 0.2, 0.3, 0.4, 0.5, 0.6\\
$f_{\rm dust}$                                 &0.01, 0.1, 0.2, 0.3\\
$E (B-V)_{\rm *}/E(B-V)_{\rm neb}$    &0.4, 0.5, 0.6, 0.7, 0.8, 1.0 \\
Dust PL Slope ($\delta$)  	               &-0.25, 0 \\
$q_{\rm PAH}$                              &0.47, 1.12, 1.77, 2.50, 3.19, 3.90, 4.58 \\
$\gamma$                                &0.01, 0.05, 0.1, 0.5 \\
$\alpha$                                   &2.0 \\
$U_{\rm min}$                                &0.3, 0.5, 0.7, 1.0 \\
\hline \hline
\end{tabular}
\caption{The electron density ($n_{\rm e}$) defines the gas density of the ionized regions and impacts the strength of nebular emission. The ionization parameter (log(U)) describes the ratio of ionizing photon density to gas density and regulates emission line intensities. The escape fraction ($f_{\rm esc}$) represents the fraction of ionizing photons that escape the system without contributing to nebular emission, while $f_{\rm dust}$ defines the fraction of Lyman continuum photons absorbed by dust.
The $E (B-V)_{*}/E(B-V)_{\rm neb}$ parametrizes the differential attenuation between stellar continuum and nebular emission. The dust attenuation power-law slope ($\delta$) modifies the shape of the attenuation curve relative to the standard prescription. The parameter $q_{\rm PAH}$ represents the fraction of dust mass in polycyclic aromatic hydrocarbons. The parameters $\gamma$, $\alpha$, and $U_{\rm min}$ are associated with the dust emission model of Draine \& Li (2007). Specifically, $\gamma$ is the fraction of dust mass exposed to a power-law distribution of radiation field intensities, $\alpha$ is the slope of that distribution, and $U_{\rm min}$ is the minimum interstellar radiation field heating the diffuse dust component. Together, these parameters regulate the shape and intensity of the infrared dust emission in the modeled SEDs. Overall, our choice of model parameters to fit the eYSC-I, eYSC-II, F335M-selected, and F770W-selected sources in NGC 628 results in a total model grid of 153538560 individual SEDs.}
\end{table*}

%% file: table_parameters_optical.tex
\begin{table*}
\centerline{SED Model Parameters for Optical Star Clusters}
\centering
\begin{tabular}{ll}
\hline
\hline
Star Formation History          & Instantaneous Burst\\
Reddening \& Extinction         & $E(B - V)=$~[0 - 1.5]~mag; $\Delta=0.05$~mag; $R_V=A_V/E(B{-}V)=3.1$\\
Age                            & 1, 2, 3, 4, 5, 6, 7, 8, 9, 10, 11, 12, 13, 14, 15, 16, 18, 19, 20, 25, 30, 35, 40, 45, 50, 55, 60  \\ 
				 & 65, 70, 75, 80, 85, 90, 95, 100, 150, 200, 250, 300, 350, 400, 450, 500, 550, 600, 650, 700  \\
				 & 750, 800, 850, 900, 950, 1000, 1500, 2000, 2500, 3000, 3500, 4000, 4500, 5000, 5500, 6000  \\
				 & 6500, 7000, 7500, 8000, 8500, 9000, 9500, 10000, 11000, 12000, 13000 \\ 			 
SSP model                       & Bruzual \& Charlot 2003 \\
Metallicity                     & $Z=0.004, 0.02, 0.05$ \\
IMF                             & Chabrier et al. 2003; [1 - 100]~$M_\odot$; fully sampled\\
$n_{\rm e}$                                 &100 \\
Log(U)                                  &-3.5, -2.0\\
$f_{\rm esc}$                                 &0.01, 0.1, 0.2, 0.4, 0.6 \\
$f_{\rm dust}$                                 &0.1, 0.2, 0.3\\
$E (B-V)_{\rm *}/E(B-V)_{\rm neb}$    &0.4, 0.5, 0.6, 0.7, 0.8, 1.0 \\
Dust PL Slope ($\delta$)   	               &-0.25, 0 \\
$q_{\rm PAH}$                              &0.47, 1.12, 1.77, 2.50, 3.19 \\
$\gamma$                                &0.01, 0.05, 0.1 \\
$\alpha$                                   &2.0 \\
$U_{\rm min}$                                &0.3, 0.5, 0.7 \\
\hline \hline
\end{tabular}
\caption{See Table 2 for a full description of each model parameter. Overall, our choice of model parameters to fit the optically-selected sources in NGC 628 results in a total model grid of 772001280 individual SEDs.}
\end{table*}